\documentclass[12pt]{article}
\usepackage{graphicx}
\usepackage{epsfig}
\usepackage{latexsym}
\usepackage{latexsym}
\usepackage{amssymb}
\usepackage{amsmath}

\begin{document}

\author{Khireddine Nouicer\thanks{
khnouicer@yahoo.fr} \\
\textit{Laboratory of Theoretical Physics and Department of Physics,}\\
\textit{\ Faculty of Sciences, University of Jijel}\\
\textit{\ Bp 98, Ouled Aissa, 18000 Jijel, Algeria.}}
\title{Regularized One Dimensional Coulomb Potential Induced by the Presence
of Minimal Lengths}
\date{}
\maketitle

\begin{abstract}
In this paper we address the problem of a particle moving in singular one
dimensional potentials in the framework of quantum mechanics with minimal
length. Using the momentum space representation we solve exactly the Schrodinger
equation for the Dirac delta potential and Coulomb potential. The effect of
the minimal length is revealed by a computation of effective generalized
Dirac delta potential and Coulomb potential.

\end{abstract}

\section{Intoduction}

In a serie of papers Kempf et al. \cite{kempf0, kempf1, kempf2} introduced a
deformed quantum mecahnics based on generalized uncertainty principle (GUP).
Other similar issues leading to the same GUP have been also initiated by
some authors \cite{hossen1, hossen2,hossen3,sastry}. One major consequence
of the GUP is the appearance of minimal uncertainties in position and/or
momentum leading to an UV/IR mixing, which allows to probe short distance
physics (UV) from long distance one (IR). A minimal uncertainty in position
or minimal length has appeared in different context like string theory \cite%
{ven,groos,amati}, loop quantum gravity \cite{Garay} and non-commutative
field theories \cite{qft1,qft2}. The minimal length defines a scale below it
the physics becomes inaccessible, leading to a natural cut-off which
prevents from the usual UV divergencies.

Recently the Schrodinger equation in momentum space for the harmonic
oscillator with minimal lengths in arbitrary dimensions has been solved \cite%
{kempf0,kempf1,minic}. The cosmological constant problem and the classical
limit of the physics with minimal lengths have been also investigated \cite%
{chang,chang01}. On the other hand the effect of the minimal length on the
properties of the 3D Coulomb potential has been studied in Refs. \cite%
{brau,akhoury,Benzik} and of the 3D Dirac oscillator using supersymmetric
quantum mechanics \cite{quesne}. The Casimir force for the electromagnetic
field in the presence of the minimal length has been also computed \cite%
{nouicer,harbach}.

In this paper we are interested by the effect of the presence of minimal
lengths on the energy eigenvalues and eigenfunctions of one dimensional
quantum mechanical systems in singular potentials, like the attractive Dirac
potential and Coulomb potential. In section 2, after implementing the minimal length via generalized position and momentum operators, 
we derive fundamental relations for our caulculations. In sections 3 and 4, the one dimensional attractive Dirac potential and Coulomb potential are considered in great details. The last section is left for concluding remarks.

\section{Quantum mechanics with minimal length}

Following \cite{sastry} we consider the following one dimensional
realization of the position and momuntum operators \
\begin{equation}
X=i\hbar \exp (P^{2}/2\mu ^{2})\frac{d}{dp}\exp (P^{2}/2\mu ^{2})\qquad
P=p,\quad \beta \geqslant 0.  \label{xp}
\end{equation}%
where $\mu $ is a parameter assumed to be large. The representation $\left( %
\ref{xp}\right) $leads to the following generalized commutator and
generalized uncertainty principle (GUP)
\begin{equation}
\left[ X,P\right] =i\hbar \exp (P^{2}/\mu ^{2}),\quad \Delta X\Delta P\geq
\frac{\hbar }{2}\left[ 1+\frac{(\Delta p)^{2}}{\mu ^{2}}\right] .
\label{gup}
\end{equation}%
A consequence of this relation, is the appearance of a minimal length given
by
\begin{equation}
\left( \Delta x\right) _{min}=\frac{\hbar }{\mu },
\end{equation}%
which means a lost of localized states in the $x$-space since we cannot
probe the coordinates space with a resolution less than the minimal length.
In this situation we use the momentum space representation and maximally
localized states to define a quasi position representation \cite{kempf1}.
Then the existence of a one parameter familly of position basis related to
the minimal length allows us to write $X\mid x^{ML}>=x\mid x^{ML}>$ where
the vectors $\mid x^{ML}>$ represent maximal localisation states in the
sence that $<X>_{ML}=x$. They are normalized states unlike the ones of
ordinary quantum mechanics. In the following we derive neccessary relations
for our calculation taking in mind that we must recover the usual quantum
mechanics in the limit $\mu \rightarrow \infty $.

Using the maximally localized vectors we derive \ the following maximally
localized wavefunction
\begin{equation}
<x^{ML}\mid p>=N\exp \left( -\frac{ix}{2\hbar }\mu \sqrt{\pi }\hbox{erf}%
\left( \frac{p}{\mu }\right) -\frac{p^{2}}{2\mu ^{2}}\right) .  \label{plane}
\end{equation}%
These states are far from being the well known plane waves. However in the limit $%
\mu \rightarrow \infty $ and choosing the normalization constant $N$ \ as $1/%
\sqrt{2\pi \hbar }$ we recover the usual planes waves of ordinary quantum
mechanics.\newline
On the other hand the states defined by Eq.(\ref{plane}) are physical ones
since \ \

\begin{equation}
<\frac{p^{2}}{2m}>=\frac{1}{4\pi m\hbar }\int p^{2}dp\exp \left( -\frac{p^{2}%
}{\mu ^{2}}\right) =\frac{\mu ^{\frac{3}{2}}}{16\sqrt{\pi }m\hbar }
\end{equation}%
Using the completness relation $1=\int_{-\infty }^{+\infty }dx\mid
x^{ML}><x^{ML}\mid $ and $\left( \ref{plane}\right) $ we obtain
\begin{equation}
<p^{\prime }\mid p>=\frac{2e^{-\frac{1}{2\mu ^{2}}\left( p^{2}+p^{\prime
2}\right) }}{\mu \sqrt{\pi }}\delta \left( \hbox{erf}\left( \frac{p}{\mu }%
\right) -\hbox{erf}\left( \frac{p^{\prime }}{\mu }\right) \right) .
\end{equation}%
With the aid of the relation $\delta f(\lambda )=\Sigma _{i}\frac{\delta
(\lambda -\lambda _{i})}{f^{\prime }(\lambda _{i})}$, where $\lambda _{i}$
are the roots of $f(\lambda )$, we finally get%
\begin{equation}
<p^{\prime }\mid p>=\delta (p-p^{\prime }).
\end{equation}%
From this equation we have the usual completeness relation for the
eigenstates $\mid p>$
\begin{equation}
\int dp\mid p><p\mid =\mathbf{{1}.}
\end{equation}%
Let us finally show that the maximally localized states $\mid x^{ML}>,$ like
the coherent states, do not form an orthogonal set. Indeed we have%
\begin{eqnarray}
<y^{ML}\mid x^{ML}> &=&\int dp<y^{ML}\mid p><p\mid x^{ML}>  \notag \\
&=&\frac{1}{2\pi \hbar }\int_{-\infty }^{+\infty }dpe^{-\frac{p^{2}}{\mu ^{2}%
}}\exp \left\{ -\frac{i(y-x)}{2\hbar }\mu \sqrt{\pi }\hbox{erf}\left( \frac{p%
}{\mu }\right) \right\}
\end{eqnarray}%
Using the variable $q$ defined by $q=\frac{\mu \sqrt{\pi }}{2}\hbox{erf}%
\left( \frac{p}{\mu }\right) $ we have

\begin{eqnarray}
&<&y^{ML}\mid x^{ML}>=\frac{1}{2\pi \hbar }\int_{-\frac{\mu \sqrt{\pi }}{2}%
}^{+\frac{\mu \sqrt{\pi }}{2}}dq\exp \left\{ -\frac{i(y-x)}{\hbar }q\right\}
\nonumber \\
&=&\frac{1}{\pi (y-x)}\sin \left( \frac{\mu \sqrt{\pi }}{2\hbar }%
(y-x)\right) .
\end{eqnarray}%
The right hand is a well behaved function unlike the Dirac distribution of
ordinary quantum mechanics. It is clear that the limit $\mu \longrightarrow
\infty $ restores the usual normalization $<x^{ML}\mid y^{ML}>=\delta \left(
x-y\right) $.

Consider now the momentum representation of the one dimensional Schr\"{o}%
dinger equation for a particle of mass $m$ in the potential $V(x),$

\begin{equation}
\left( \frac{P^{2}}{2m}+V(X)\right) \mid \Psi >=E\mid \Psi >.
\end{equation}%
Let us project on the momentum states $\mid p>$ and inserting the closure
relations for the maximally localized states and the momentum states
respectivelly we have
\begin{equation}
\frac{p^{2}}{2m}\Psi (p)+\int_{-\infty }^{+\infty }dp^{\prime }\int_{-\infty
}^{+\infty }dx<p\mid V(X)\mid x^{ML}><x^{ML}\mid p^{\prime }><p^{\prime
}\mid \Psi >=E\Psi (p).
\end{equation}%
Then using the expression of the quasi-position eigenvectors we obtain the following equation

\begin{equation}
\frac{p^{2}}{2m}\Psi (p)+\frac{1}{2\pi \hbar }\int_{-\infty }^{+\infty
}dp^{\prime }e^{-\frac{1}{2\mu ^{2}}\left( p^{2}+p^{\prime 2}\right)
}\int_{-\infty }^{+\infty }dxe^{-\frac{i\lambda \mu }{2\hbar }\sqrt{\pi }%
\left( \hbox{erf}\left( \frac{p}{\mu ^{2}}\right) -\hbox{erf}\left( \frac{%
p^{\prime }}{\mu ^{2}}\right) \right) }V(x)\Psi (p^{\prime })=E\Psi (p),
\end{equation}%
which can be written as

\begin{equation}
\frac{p^{2}}{2m}\Psi (p)+\int_{-\infty }^{+\infty }dp^{\prime }e^{-\frac{1}{%
2\mu ^{2}}\left( p^{2}+p^{\prime 2}\right) }V(p,p^{\prime })\Psi (p^{\prime
})=E\Psi (p),  \label{schro}
\end{equation}%
where the potential $V(p,p^{\prime })$ is the generalized Fourier transform
of the potential $V(x)$
\begin{equation}
V(p,p^{\prime })=\frac{1}{2\pi \hbar }\int_{-\infty }^{+\infty }e^{-\frac{%
i\lambda \mu }{2\hbar }\sqrt{\pi }\left( \hbox{erf}\left( \frac{p}{\mu ^{2}}%
\right) -\hbox{erf}\left( \frac{p^{\prime }}{\mu ^{2}}\right) \right)
}V(x)dx.  \label{fourier}
\end{equation}

\section{One dimensional delta potential}

Let us begin our investigation of the effect of minimal length on
eigenfunctions and eigenenergies by considering the one dimensional Dirac
delta potential $V(x)=-\kappa \delta \left( x\right) .$ This potential is
known to shares some common properties with the Coulomb potential in the
limit of strong coupling. Using the generalized Fourier transform $\left( %
\ref{fourier}\right) $ we obtain the usual expression

\begin{equation}
V(p-p^{\prime })=-\frac{\kappa }{2\pi \hbar }.
\end{equation}%
Substituting in Schr\"{o}dinger equation $\left( \ref{schro}\right) $ we have

\begin{equation}
\left( \frac{p^{2}}{2m}-E\right) \Psi \left( p\right) -\frac{\kappa }{2\pi
\hbar }e^{-\frac{1}{2\mu ^{2}}p^{2}}\int_{-\infty }^{+\infty }dp^{\prime
}e^{-\frac{1}{2\mu ^{2}}p^{\prime 2}}\Psi \left( p^{\prime }\right) =0.
\end{equation}%
This equation can be written in the following integral form

\begin{equation}
\Psi \left( p\right) =\lambda \int_{-\infty }^{+\infty }dp^{\prime
}K(p,p^{\prime })\Psi \left( p^{\prime }\right)  \label{wav}
\end{equation}%
with $\lambda =1$ and $K(p,p^{\prime })$ a separable kernel given by

\begin{equation}
K(p,p^{\prime })=f(p)g(p^{\prime }),
\end{equation}%
where

\begin{equation}
f(p)=\frac{m\kappa }{\pi \hbar }\frac{e^{-\frac{1}{2\mu ^{2}}p^{2}}}{\left(
p^{2}-2mE\right) },\text{\quad\ }g(p)=e^{-\frac{1}{2\mu ^{2}}p^{\prime 2}}.
\label{fg}
\end{equation}%
Writing the wave function in the following form%
\begin{equation}
\Psi \left( p\right) =cf(p)\quad \text{with \quad }c=\int_{-\infty
}^{+\infty }dp^{\prime }g(p^{\prime })\Psi \left( p^{\prime }\right) ,
\label{wave}
\end{equation}%
we deduce the equation

\begin{equation}
\Psi \left( p\right) \left[ 1-\int_{-\infty }^{+\infty }dpg(p)f\left(
p\right) \right] =0
\end{equation}%
which admits the obvious solution

\begin{equation}
\int_{-\infty }^{+\infty }dpg(p)f\left( p\right) =1.
\end{equation}%
This equation is the spectral condition from which we extract the possible
energy levels. Using the expressions of $f(p)$ and $g(p)$ and performing the
integral we obtain

\begin{equation}
\frac{m\kappa }{p_{0}\hbar }\left[ 1-\hbox{erf}\left( \frac{p_{0}}{\mu }%
\right) \right] e^{\frac{p_{0}^{2}}{\mu ^{2}}}-1=0,  \label{int}
\end{equation}%
where we set $2mE$ =-$p_{0}^{2}$ for bound states$.$ An asymptotic expansion
in $\frac{1}{\mu }$ gives the following equation

\begin{equation}
\left( 1+\frac{2m\varkappa }{\hbar \sqrt{\pi }\mu }\right) p_{0}-\frac{%
m\varkappa }{\hbar }=0.  \label{m01}
\end{equation}%
Using the fact that $\frac{1}{\mu }$ is a small parameter, the solution of $%
\left( \ref{m01}\right) $ gives the following energy eigenvalue

\begin{equation}
E=-\frac{m\kappa ^{2}}{2\hbar ^{2}}\left( 1-\frac{2m\kappa }{\hbar \sqrt{\pi
}\mu }\right) ^{2}.
\end{equation}%
An asymptotic expansion in $\mu $ gives

\begin{equation}
E=-\frac{m\kappa ^{2}}{2\hbar ^{2}}\left( 1-4\frac{m\kappa }{\hbar \sqrt{\pi
}\mu }+4\left( \frac{m\kappa }{\hbar \sqrt{\pi }\mu }\right) ^{2}\right) .
\label{energy0}
\end{equation}%
Using the normalization condition $\int dp\left\vert \Psi \left( p\right)
\right\vert ^{2}=1$ and the the following formula \cite{grad}

\begin{eqnarray}
\int_{0}^{\infty }y^{\nu -1}\left( y+\gamma \right) ^{\lambda -1}e^{-\frac{%
\beta }{y}}dy &=&\beta ^{\frac{\nu -1}{2}}\gamma ^{\frac{\nu -1}{2}+\lambda
}\Gamma \left( 1-\nu -\lambda \right) e^{\frac{\beta }{2\gamma }}W_{\frac{%
\nu -1}{2}+\lambda ,-\frac{\nu }{2}}\left( \frac{\beta }{\gamma }\right) , \\
\left\vert \arg \gamma \right\vert &<&\pi ,\quad \hbox{Re}\left( 1-\lambda
\right) >\hbox{Re}\nu >0.  \notag
\end{eqnarray}%
to calculate the constant $c$, the normalized wave function in momentum space is finally given by

\begin{equation}
\Psi \left( p\right) =\left[ \mu ^{-1/2}p_{0}^{-5/2}\sqrt{\pi }W_{-\frac{3}{4%
},-\frac{3}{4}}\left( \frac{p_{0}^{2}}{\mu ^{2}}\right) \right] ^{-1/2}\frac{%
e^{-\frac{p_{0}^{2}}{4\mu ^{2}}}e^{-\frac{1}{2\mu ^{2}}p^{2}}}{\left(
p^{2}+p_{0}^{2}\right) }.
\end{equation}%
The maximally localized wave function $\Phi \left( x\right) $ are obtained
from the generalized inverse Fourier transform of the momentum space wave
function given by

\begin{equation}
\Phi \left( x\right) =\frac{1}{\sqrt{2\pi \hbar }}\int_{-\infty }^{+\infty
}dp\exp \left( \frac{ix}{2\hbar }\mu \sqrt{\pi }\hbox{erf}\left( \frac{p}{%
\mu }\right) -\frac{p^{2}}{2\mu ^{2}}\right) \Psi \left( p\right) .
\end{equation}%
Thus using eq.(28) we obtain

\begin{equation}
\Phi \left( x\right) =\frac{1}{\sqrt{2\pi \hbar }}\left[ \mu
^{-1/2}p_{0}^{-5/2}\sqrt{\pi }W_{-\frac{3}{4},-\frac{3}{4}}\left( \frac{%
p_{0}^{2}}{\mu ^{2}}\right) \right] ^{-1/2}e^{-\frac{p_{0}^{2}}{4\mu ^{2}}%
}\int_{-\infty }^{+\infty }dpe^{-\frac{p^{2}}{\mu ^{2}}}\frac{\exp \left(
\frac{ix}{2\hbar }\mu \sqrt{\pi }\hbox{erf}\left( \frac{p}{\mu }\right)
\right) }{\left( p^{2}+p_{0}^{2}\right) }.
\end{equation}%
Using the residus theorem we show that the second integral vanishes and we
obtain

\begin{equation}
\Phi \left( x\right) =\left[ 2\hbar \mu ^{-1/2}p_{0}^{-9/2}\pi ^{-5/2}W_{-%
\frac{3}{4},-\frac{3}{4}}\left( \frac{p_{0}^{2}}{\mu ^{2}}\right) \right]
^{-1/2}e^{\frac{3p_{0}^{2}}{4\mu ^{2}}}\left[ e^{-\frac{\left\vert
x\right\vert }{2\hbar }\mu \sqrt{\pi }\text{erfi}\left( \frac{p_{0}}{\mu }%
\right) }\right] ,  \label{c}
\end{equation}%
where erfi$\left( x\right) =-i\hbox{erf}(ix)$ is the imaginary error
function.

Let us now define $\Phi \left( p\right) =e^{\frac{p^{2}}{2\mu ^{2}}}\Psi
\left( p\right) $ and rewrite Schr\"{o}dinger equation with an effective
potential, in the manner of the theory without the minimal length,

\begin{equation}
\left( \frac{p^{2}}{2m}-E\right) \Phi \left( p\right) +\int_{-\infty
}^{+\infty }dp^{\prime }V_{\text{eff}}\left( p,p^{\prime }\right) \Phi
(p^{\prime })=0,
\end{equation}%
where

\begin{equation}
V_{\text{eff}}\left( p\right) =-\frac{\kappa }{2\pi \hbar }e^{-\frac{p^{2}}{%
\mu ^{2}}}.
\end{equation}%
Using the inverse Fourier transform we obtain the following regularized
Dirac potential in one dimension (see Figure 1.)

\begin{equation}
V_{\text{eff}}\left( x\right) =-\frac{\kappa \mu \sqrt{\pi }}{2\pi \hbar }%
e^{-\left( \frac{\mu x}{2\hbar }\right) ^{2}}.  \label{effect}
\end{equation}

\begin{figure}
\begin{center}
\includegraphics[height=08cm, width=08cm]{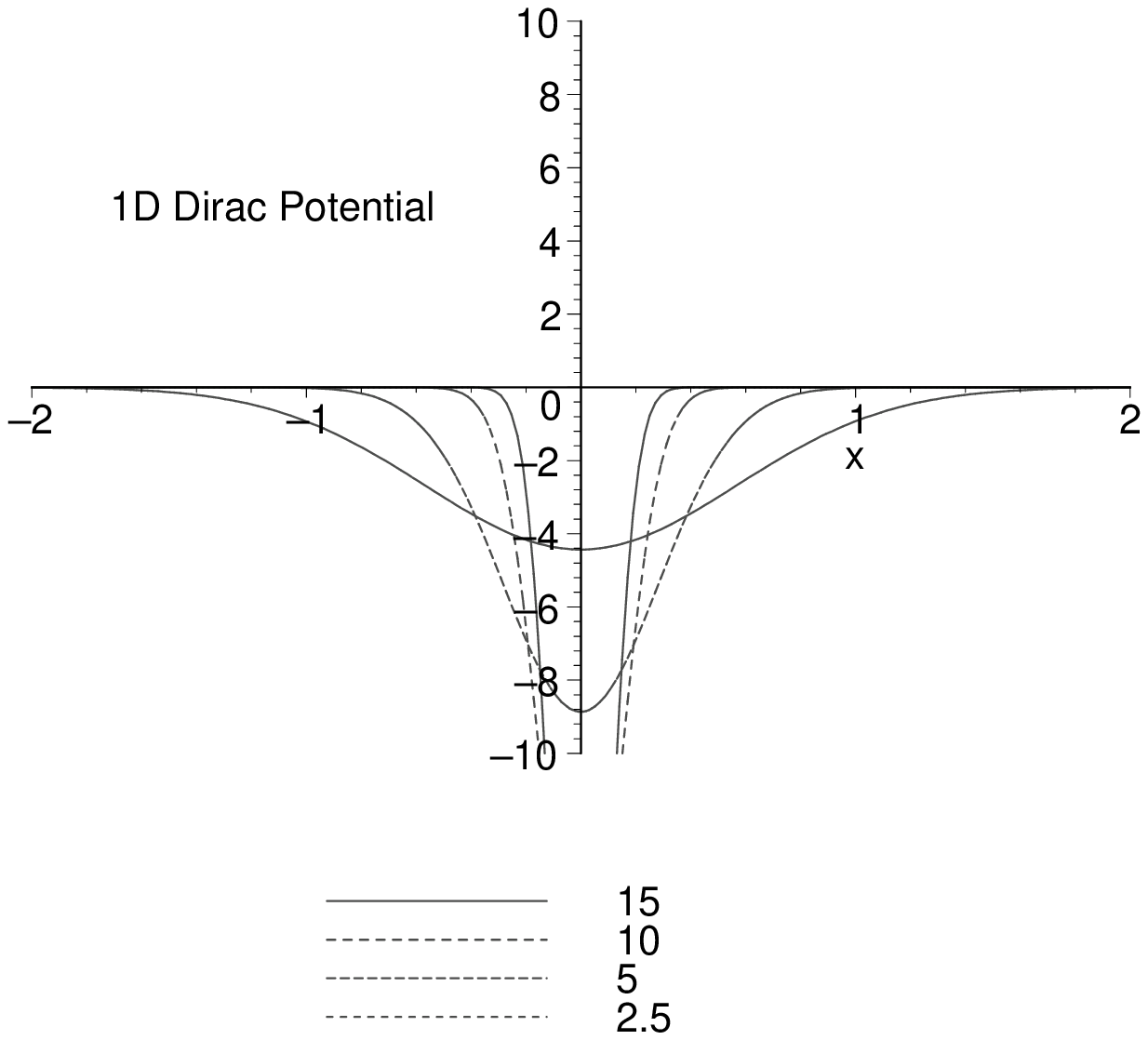}
\end{center}
\caption{Figure 1: Effective
one dimensional Dirac potential for different values of the
minimal length.}
\end{figure}

The results in the case without the minimal length are obtained by
observing, that for large $\mu ,$ we have \cite{grad}

\begin{equation}
W_{-\frac{3}{4},-\frac{3}{4}}\left( \frac{p_{0}^{2}}{\mu ^{2}}\right)
\underset{\mu \rightarrow \infty }{\rightarrow }\left( \frac{\pi \mu }{4p_{0}%
}\right) ^{\frac{1}{2}},
\end{equation}%
and then the eigenfunctions $\Psi \left( p\right) $ and $\Phi \left(
x\right) $ reduce respectively to the momentum and coordinate sapce
eigenfunctions of the one dimentional Dirac potantial without the minimal
length.

\begin{equation}
\Psi \left( p\right) =\frac{\frac{m\kappa }{\hbar }\sqrt{\frac{2m\kappa }{%
\pi \hbar }}}{\left( p^{2}+\left( \frac{m\kappa }{\hbar }\right) ^{2}\right)
},\quad \Phi \left( x\right) =\sqrt{\frac{m\kappa }{\hbar ^{2}}}e^{-\frac{%
m\kappa }{\hbar ^{2}}\left\vert x\right\vert }.
\end{equation}

\section{One dimensional coulomb potential}

Finally we consider the one dimensional coulomb potential $V(X)=-\frac{Ze^{2}%
}{X}$ in the presence of a minimal length. This potential without the
minimal length, which seems to have applications in semiconductors or
insulators, has been solved with different results \cite{reyes,ran}. In the
presence of the minimal length of Kempf et al. \cite{kempf1}, this potential
has been recently solved but with some incorrectness \cite{chuk}.

Using the following formula

\begin{equation}
\lim_{\epsilon \rightarrow 0}\int_{-\infty }^{+\infty }\frac{e^{-ipz}}{z\pm
i\epsilon }dz=\text{\textbf{P}}\int_{-\infty }^{+\infty }\frac{e^{-ipz}}{z}%
dz\mp i\pi ,
\end{equation}%
where \textbf{P} means the Cauchy principal value of the integral, the
Coulomb potential in momentum space is then given by%
\begin{equation}
V(p,p^{\prime })=\frac{iZe^{2}}{2\hbar }\left[ \hbox{sign}\left( \hbox{erf}%
\left( p\sqrt{\beta }\right) -\hbox{erf}\left( p^{\prime }\sqrt{\beta }%
\right) \right) +1\right] .
\end{equation}%
Substituting in Schrodinger's equation (\ref{schro}), we obtain
\begin{equation}
\left( \frac{p^{2}}{2m}-E\right) \Psi (p)+\frac{iZe^{2}}{\hbar }e^{-\frac{1}{%
2\mu ^{2}}p^{2}}\left[ \int_{-\infty }^{p}dp^{\prime }e^{-\frac{1}{2\mu ^{2}}%
p^{\prime 2}}+\int_{-\infty }^{+\infty }dp^{\prime }e^{-\frac{1}{2\mu ^{2}}%
p^{\prime 2}}\right] \Psi (p^{\prime })=0,  \label{c1}
\end{equation}%
where the last term $\mathbf{S}=\int_{-\infty }^{+\infty }dp^{\prime }e^{-%
\frac{1}{2\mu ^{2}}p^{\prime 2}}\Psi (p^{\prime })$ is the signature of the
singularity of Coulomb potential in one dimension in coordinates space.

Deriving $\left( \ref{c1}\right) $with respect to $p$ gives%
\begin{equation}
\Psi ^{\prime }(p)+\frac{2p}{(p^{2}-2mE)}\Psi (p)+\frac{p}{\mu ^{2}}\Psi (p)+%
\frac{2miZe^{2}}{\hbar }\frac{e^{-\frac{1}{\mu ^{2}}p^{2}}}{(p^{2}-2mE)}\Psi
(p)=0.  \label{c2}
\end{equation}%
The solution to this equation is given by

\begin{equation}
\Psi _{n}\left( p\right) =N\frac{e^{-\frac{p^{2}}{2\mu ^{2}}}}{%
(p^{2}+p_{0}^{2})}\exp \left( -\frac{2miZe^{2}}{\hbar }\int \frac{e^{-\frac{1%
}{\mu ^{2}}p^{2}}}{(p^{2}+p_{0}^{2})}dp\right) .
\end{equation}%
with the normalization constant, calculated in section 3, given by

\begin{equation}
N=\left[ \mu ^{-1/2}p_{0}^{-5/2}\sqrt{\pi }e^{\frac{p_{0}^{2}}{2\mu ^{2}}%
}W_{-\frac{3}{4},-\frac{3}{4}}\left( \frac{p_{0}^{2}}{\mu ^{2}}\right) %
\right] ^{-1/2}.
\end{equation}%
Expanding the last term in $%
1/\mu ^{2}$ and integrating over $p$ we obtain

\begin{equation}
\Psi _{n}\left( p\right) =N\frac{\exp \left( -\frac{p^{2}}{2\mu ^{2}}+\frac{%
2miZe^{2}}{\hbar \mu ^{2}}p\right) }{(p^{2}+p_{0}^{2})}\exp \left( -\frac{%
2miZe^{2}}{\hbar p_{0}}\left( 1+\frac{p_{0}^{2}}{\mu ^{2}}\right) \arctan
\frac{p}{p_{0}}\right) .  \label{c3}
\end{equation}%
The requirement that the wave function must be single valued gives the
spectral condition

\begin{equation}
\left( \frac{p_{0}^{2}}{\mu ^{2}}-\frac{\hbar n}{mZe^{2}}p_{0}+1\right) =0.
\end{equation}%
Solving this equation and using $E_{n}=-\frac{p_{0}^{2}}{2m}$ we obatin

\begin{equation}
E_{n}^{\pm }=-\frac{\mu ^{2}}{8m}\left( \frac{\hbar n\mu }{mZe^{2}}\right)
^{2}\left( 1\pm \sqrt{1-\left( \frac{2mZe^{2}}{\hbar n\mu }\right) ^{2}}%
\right) ^{2}
\end{equation}%
An asymtotic expansion in $\mu $ gives

\begin{equation}
E_{n}^{-}=-\frac{mZ^{2}e^{4}}{2\hbar ^{2}n^{2}}\left( 1-2\left( \frac{mZe^{2}%
}{\hbar n\mu }\right) ^{2}\right)  \label{mr}
\end{equation}%
and%
\begin{equation}
E_{n}^{+}=-\frac{\mu ^{2}}{2m}\left( -2+\left( \frac{\hbar n\mu }{mZe^{2}}%
\right) ^{2}-\left( \frac{mZe^{2}}{\hbar n\mu }\right) ^{2}\right) .
\label{mr2}
\end{equation}%
The first term in $\left( \ref{mr}\right) $ is the spectrum of the usual one
dimensional Coulomb potential \cite{ran}, while the remaining ones are the
corrections due to the perturbation of the space by the presence of the
minimal length. The correction term in $\left( \ref{mr}\right) $ is in
concordance with the ones to the energy spectrum of Coulomb potential in
three dimensions derived in \cite{brau,akhoury}. In \cite{chuk} an
additional correction term proportional to $\sqrt{\beta }$ $\left(\frac{1}{\mu } \text{ }%
\text{ in the paper}\right) $ \ has been taken into account.
However as noted in \cite{akhoury}, this term like e$^{\frac{2miZe^{2}}{%
\hbar \mu ^{2}}p}$ in eq.$\left( \ref{wave}\right) ,$ will affect the
outgoing scattered waves.

The wave functions associated with the energy levels given by $\left( \ref%
{mr2}\right) $ are physically acceptable ones unless $\mu $ is large but
finite. In the two cases we can extract a bound for the minimal length. In
fact requiring that $E_{n}^{\mp}<0$ for bound states we obtain

\begin{equation}
\left( \Delta x\right) _{\text{min}}\leq \frac{\hbar ^{2}n}{\sqrt{2}Ze^{2}m}.
\end{equation}%
Setting $n=1$ and the fact that Bohr radius is $5.292\times 10^{-11}$ m we
obtain

\begin{equation}
\left( \Delta x\right) _{\text{min}}\leq 3.742\times 10^{-11}\text{ m.}
\end{equation}%
Let us finally discuss about the original singularity of Coulomb potential
in one dimension expressed by $\mathbf{S}=\int_{-\infty }^{+\infty
}dp^{\prime }e^{-\frac{1}{2\mu ^{2}}p^{\prime 2}}\Psi (p^{\prime })$ . As
noted above, $\mathbf{S}$ \ is\ an account of the discontinuity at the
origin of the wave function in the coordinates space$.$With the aid of the
wavefunctions given by $\left( \ref{c3}\right) $ we show that

\begin{eqnarray}
\mathbf{S} &=&\lim_{p\rightarrow \infty }\frac{i\hbar e^{\frac{1}{2\mu ^{2}}%
p^{2}}}{2mZe^{2}}(p^{2}+p_{0}^{2})\Psi _{n}(p)  \notag \\
&\rightarrow &_{p\rightarrow \infty }\exp \left( \frac{2miZe^{2}}{\hbar \mu
^{2}}p\right) .  \label{s}
\end{eqnarray}%
As it is obvious from $\left( \ref{s}\right) ,$ the singularity is
completely removed by the presence of the minimal length. This is a natural
consequence since the minimal length play the role of a cut-off which
suppress the contribution of high momentum. This was not the situation in
\cite{chuk}.

The regularizing effect of the minimal length is best seen by computing the
effective Coulomb potential induced by the presence of the minimal. Indeed
reappeting the same calculations leading to $\left( \ref{effect}\right) $
along with $\mathbf{S}=0$ and taking the inverse Fourier sine transform we
obtain

\begin{equation}
V_{\text{eff}}(x)=-\frac{Ze^{2}\sqrt{\pi }}{2\hbar }\mu \text{erfi}\left(
\frac{\mu x}{2}\right) e^{-\left( \frac{\mu x}{2\hbar }\right) ^{2}}.
\label{eff}
\end{equation}%
We note that the usual divergence is removed and the presence of the minimal
length regularizes the theory (see Figure 2 below). Similar effective
Coulomb potential in three dimensions induced by the non commutativity of
the space-time coordinates have been recently computed \cite{grupuso}

\begin{figure}
\begin{center}
\includegraphics[height=10.0364cm, width=10.0364cm]{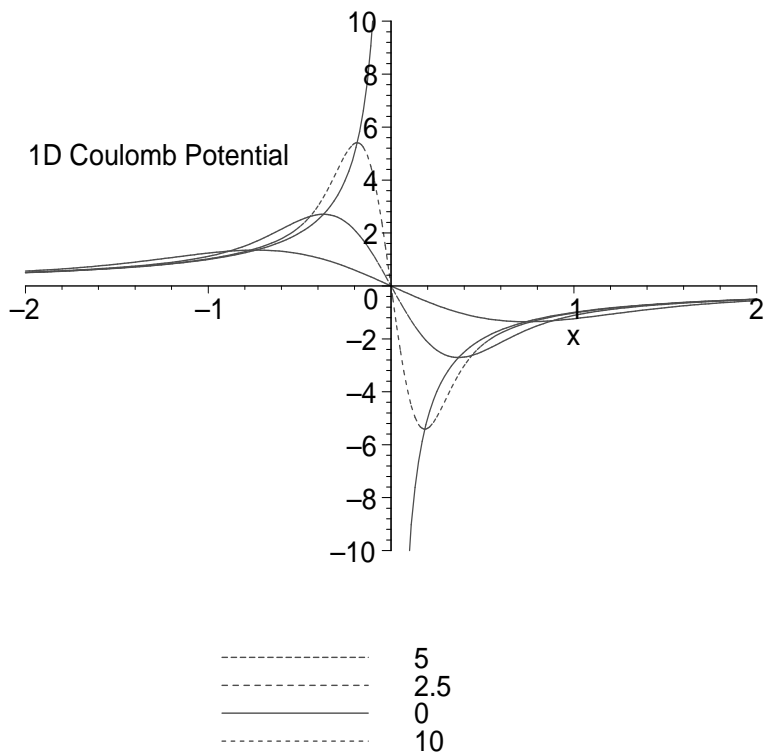}
\end{center}
\caption{Figure 2: Effective
one dimensional Coulomb potential for different values of the
minimal length.}

\end{figure}

\section{Conclusion}

In this paper we have solved, in the framework of quantum mechanics in the
presence of minimal lengths, the problem of a particle moving in simple
singular one dimensional potentials namely, the Dirac delta potential and
the Coulomb potential. Our calculations are based on using physical maximal
localisation states. In the two cases We have calculated the effective
potential induced by the presence of the minimal length and as a consequence
we have shown that the singularity at the origin, present in the case
without the minimal length, is now completely removed. For a particle in the
Coulomb potential the corrections to the energy spectrum are in concordance
with that of Refs.\cite{brau,akhoury}.

\end{document}